\documentclass[onecolumn, amsmath,amssymb, aps,12pt]{revtex4}
\usepackage{setspace}
\usepackage[dvipsnames]{xcolor}
\usepackage{graphicx}
\usepackage{dcolumn}
\usepackage{bm}

\begin{document}

\title{Nonreciprocal Response of Electrically Biased Graphene-Coated Fiber }

\author{Asma Fallah\textsuperscript{1}}
 
 \author{Nader Engheta\textsuperscript{1}}%
 \email{engheta@seas.upenn.edu}
\affiliation{%
 \textsuperscript{1} University of Pennsylvania, Department of Electrical and Systems Engineering, Philadelphia, Pennsylvania 19104, USA
}%
\date{\today}

\newpage
\begin{abstract}
Here, we theoretically investigate the nonreciprocal response of an electrically biased graphene-coated dielectric fiber. By electrically biasing the graphene coating along the fiber axis, the dynamic conductivity of graphene exhibits a nonsymmetric response with respect to the longitudinal component of guided-mode wave vectors.  Consequently, the guided waves propagating in two opposite directions may encounter distinct propagation features. In this work, the electromagnetic properties, such as modal dispersion and some field distributions are presented, and the strength of nonreciprocity is discussed for different parameters of graphene, such as its chemical potential and material loss. Furthermore, the influence of the radius of the fiber on the nonreciprocal response is investigated. We envision that such nonreciprocal optical fibers may find various potential applications in THz, microwave, and optical technologies.
\end{abstract}

\maketitle
 \newpage
 Graphene, a two-dimensional (2D) one-atom-thick material formed by 2D arrays of carbon atoms, exhibits interesting properties in its electronic and photonic interactions, offering exciting potential applications in $\mathrm{THz}$ and optics\cite{vakil2011transformation,brar2015electronic,thongrattanasiri2012complete,esquius2014sinusoidally,wu2021graphene,wang2019graphene, liu2011graphene,shaukat2020drag}. Single-layer graphene has ultrahigh electron mobility with drift velocity that can be as high as $v_\emph{f}=c/300$ ($v_{\emph{f}}$ is Fermi velocity and $c$ is the speed of light in vacuum) \cite{mariodrift}. It has been shown that such DC drift electric current, generated by the DC biasing of a graphene sheet, can break the Lorentz reciprocity for the surface waves along the graphene, providing a significant nonreciprocal response\cite{mariodrift,morgado2017negative,grapheneborgnia,correas2015electrically,zhao2021efficient,dong2021fizeau,hassani2022drifting,morgado2021active,morgado2022directional}. The degree of nonreciprocity is influenced by the speed at which the electrons move ($v_\emph{0}$) and by their density ($n_\emph{0}$). It has been shown that significant nonreciprocity can be achieved by the mere presence of an electron beam comprising a substantial number
of electrons moving with a constant velocity \cite{fallah2021nonreciprocal,fallah2022electron} in vacuum.

Unbiased graphene-coated deeply sub-wavelength fiber has been investigated in the past, exhibiting interesting features such as wave guidance along fibers with significantly tiny radius\cite{davoyan2016salient}. Here, we consider a dielectric fiber with a graphene coating that is electrically biased with a DC voltage along the fiber axis, and we undertake a theoretical investigation to examine the nonreciprocity of various guided modes in the fiber, resulted from the effects of such electrical bias. We study the strength of nonreciprocity for three types of guided modes; transverse magnetic ($TM$), hybrid ($HE$), and surface plasmon polariton ($SPP$) modes, by presenting their modal dispersion. In addition, we have presented a field distribution for $SPP$ mode which clearly show nonreciprocal response. Moreover, the effect of the radius of the fiber is also investigated here. Furthermore, we discuss the influence of the material loss of the graphene coating on the nonreciprocal response and present how it affects the real and imaginary parts of the guided-mode wavenumbers. In this work, time variation is assumed to be $e^{i\omega t}$ with $\omega$ being the angular frequency. For simplicity, we neglect the intrinsic nonlocal response of the graphene.

\begin{figure}[htbp]
\centering\includegraphics[width=\columnwidth]{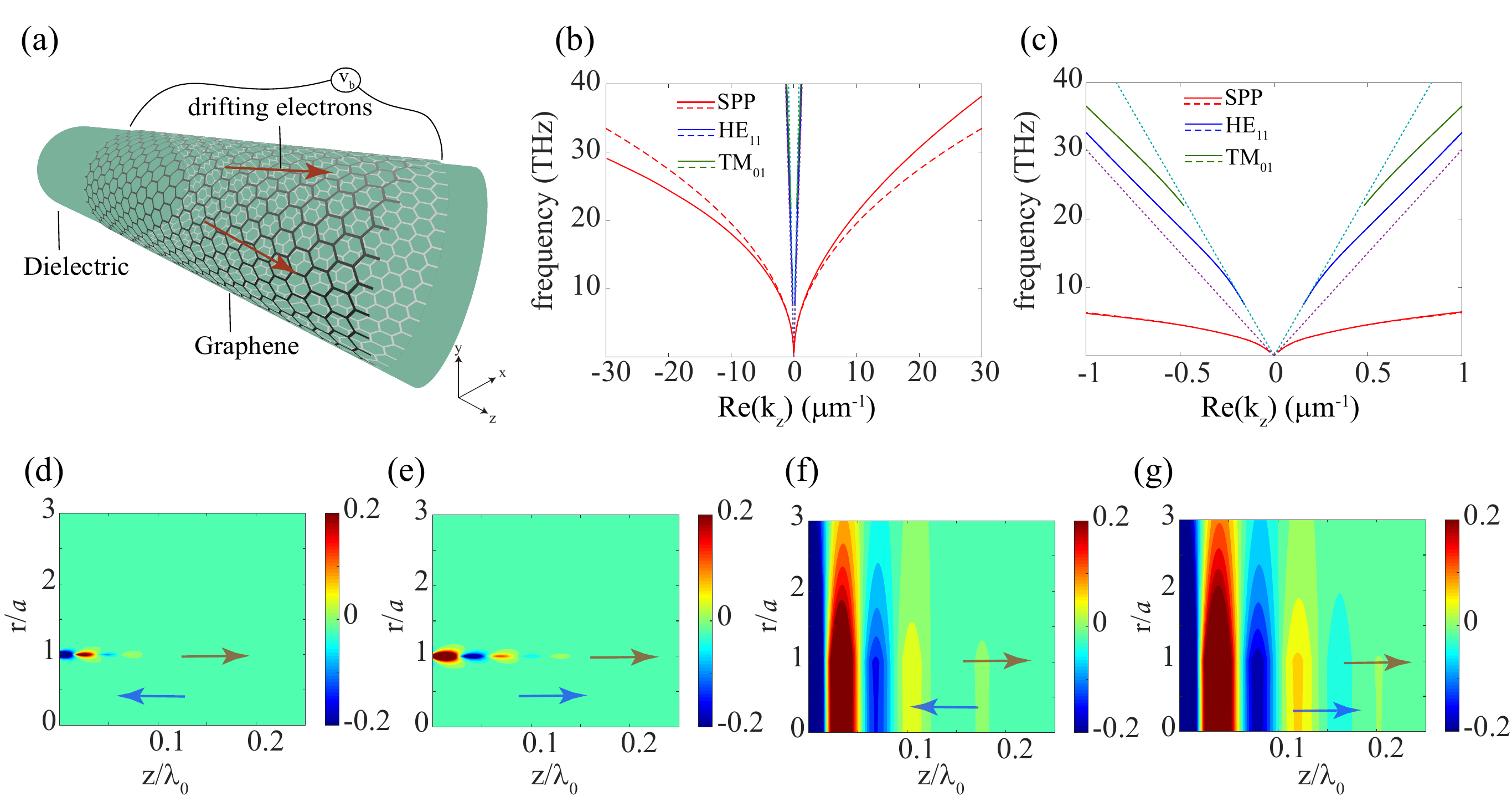}
\caption{(a) Schematic of a dielectric fiber coated with a single layer of graphene. $V_\emph{b}$ denotes the DC bias voltage alongside of the graphene coating. The red arrows conceptually represent the direction of electron drift, depending on the polarity of $V_\emph{b}$. (b) Dispersion graphs of the fiber modes, i.e., $HE_{\emph{11}}$, $TM_{\emph{10}}$, and $SPP$ modes for $v_\emph{0}=0$ and $v_\emph{0}=v_\emph{f}$ shown in dashed and solid lines, respectively. (c) zoomed-in plot around the origin of the panel (b). Dashed cyan and purple lines in (b) and (c) correspond to light lines in the air and fiber. Relative permittivity of the core is $\varepsilon_{\emph{core}}=2.5$ and its radius is $a=5$ $\mu m$. (d-g) Time snapshot of the $z$-component of the normalized electric field distribution of the guided $SPP$ mode for $f=15$ $\mathrm{THz}$ and radius $a=5$ $\mu m$ (d) and (e) and $a=0.1$ $\mu m$ (e) and (f). (d) and (f) are for $v_\emph{0}=-v_\emph{f}$ (direction of drift electrons is anti-parallel to the direction of mode propagation). (e) and (g) are for $v_\emph{0}=v_\emph{f}$ (direction of drift electrons is parallel to the direction of mode propagation). Red and blue arrows shows the direction of drift electrons and mode propagation, respectively.}
\label{fig:geo}
\end{figure}

To begin, we point out that the non-biased graphene is characterized by surface conductivity ($\sigma_{\emph{g}}$ is scalar) described by Kubo's formula \cite{hanson2008dyadic, gusynin2006magneto}. It has been demonstrated in \cite{morgado2017negative} that the dynamic conductivity of biased graphene can be written by considering Galilean Doppler shift model, meaning the electrons in the biased graphene moves as a single body with a constant velocity. In the present work, the drift velocity of electrons is assumed to be along the $z$-axis, which is the axis of the fiber. As a result, electromagnetic waves with wave vectors having $z$-component ($k_\emph{z}$) parallel or antiparallel with respect to the direction of electron drift velocity in graphene exhibit different interactions with the graphene, since the dynamic surface conductivity adopts two distinct values.

 The geometry of the problem of interest is shown in Fig. \ref{fig:geo}(a), where a dielectric fiber (shown in green) is coated with a single layer of graphene. Here, we employ Maxwell's equations in this guided-wave structure, in analogy with the modal analysis in conventional optical fibers \cite{okamoto2021fundamentals}. By incorporating the appropriate boundary conditions and considering the dynamic conductivity of graphene, we analytically derive dispersion relations for various modes: $HE_{\emph{11}}$, $TM_{\emph{10}}$, and surface plasmon polaritons ($SPP$), in this graphene-coated optical fiber with a radius of '\textit{a}' and a core relative permittivity of $\varepsilon_{\emph{core}}$. For the $HE_{\emph{11}}$ mode, denoting $k_{\emph{0}}^2 \equiv \omega^2 \mu_{\emph{0}} \varepsilon_{\emph{0}}$ , $ k_{\emph{rcore}}^2 \equiv k_{\emph{0}}^2 \varepsilon_{\emph{core}}-k_{\emph{z}}^2$, and $ k_{\emph{r0}}^2 \equiv k_{\emph{z}}^2-k_{\emph{0}}^2$, the dispersion relation is given by,

 \begin{equation}
 \begin{split}
\\
&
\frac{k_\emph{z}^2n^2}{a^2\mu_0 \omega}(\frac{1}{k_\emph{{r0}}^2}+\frac{1}{k_\emph{{rcore}}^2})^2 X Y-\varepsilon_0 \omega (\frac{\varepsilon_{\emph{core}}}{k_\emph{{rcore}}^2}\frac{Y}{X}+\frac{1}{k_\emph{{r0}}^2}\frac{X}{Y}+\frac{1+\varepsilon_{\emph{core}}}{k_\emph{{r0}}k_\emph{{rcore}}})=
\\
&
\frac{\mu_\emph{0}\omega \Tilde{\sigma}_{\varphi}  \Tilde{\sigma}_{z}}{k_\emph{{r0}}k_\emph{{rcore}}}+i \Tilde{\sigma}_{z} (\frac{X}{k_\emph{{r0}}}+\frac{Y}{k_\emph{{rcore}}})+i \frac{\Tilde{\sigma}_{\varphi} }{k_\emph{{r0}}k_\emph{{rcore}}}
  (\frac{k_z^2n^2}{a^2}(\frac{Y}{k_\emph{{r0}}^3}+\frac{X}{k_\emph{{rcore}}^3})-\varepsilon_\emph{0}\mu_\emph{0}\omega^2(\frac{\varepsilon_{\emph{core}}}{k_\emph{{rcore}}X}+\frac{1}{k_\emph{{r0}}Y}))
    \end{split}
\label{eq:drhe}
\end{equation}

where $X\equiv J_n(k_{\emph{rcore}} a)/J_n^{\prime}(k_{\emph{rcore}} a)$, $Y\equiv K_n(k_{\emph{r0}} a)/K_n^{\prime}(k_{\emph{r0}} a)$, and $n$ is an integer number, which here is assumed to be $1$. In addition, $K_\emph{n}$, $J_\emph{n}$, $\mu_\emph{0}$, and $\varepsilon_\emph{0}$ are modified Bessel functions of the second kind, Bessel function of the first kind ($X^{\prime}$ denotes derivation of the function X with respect to its argument), permeability and permittivity in free space, respectively. It is noteworthy that, despite the absence of consideration for nonlocality in this study and the assumption of equal longitudinal (${\sigma}_{z}$) and transverse (${\sigma}_{\varphi}$) conductivity in unbiased graphene, the presence of drift electrons yields disparate values for these conductivity components which are $ \tilde{\sigma}_{\emph{z}}=\left({\omega}/{\tilde{\omega}}\right)\sigma_{\emph{g}}(\tilde{\omega})$ and $ \tilde{\sigma}_{\varphi}= \left(\tilde{\omega}/{{\omega}}\right)\sigma_{\emph{g}}(\tilde{\omega})$, where $\tilde{\omega}\equiv(\omega - v_{\emph{0}} k_{\emph{z}})$ represents the Doppler-shifted frequency, and $v_{\emph{0}} k_{\emph{z}}$ is the inner product of the electron drift velocity with the wave vector of the propagating electromagnetic wave. Furthermore, for the $TM_{\emph{10}}$ mode with $E_\emph{z}$, $E_\emph{r}$, and $H_\varphi$ field components, the dispersion relation is given by,

\begin{equation}
\frac{1}{k_{\emph{r0}}} \frac{K_\emph{1}\left(k_{\emph{r0}} a\right)}{K_\emph{0}\left(k_{\emph{r0}} a\right)}+\frac{\varepsilon_{\emph{rcore}}}{k_{\emph{rcore}}} \frac{J_\emph{1}\left(k_{\emph{rcore}} a\right)}{J_\emph{0}\left(k_{\emph{rcore}} a\right)}=\frac{i \Tilde{\sigma}_{z}}{\omega \varepsilon_{\emph{0}}}.
\label{eq:drtm}
\end{equation}
\begin{figure}
\centering\includegraphics[width=\columnwidth]{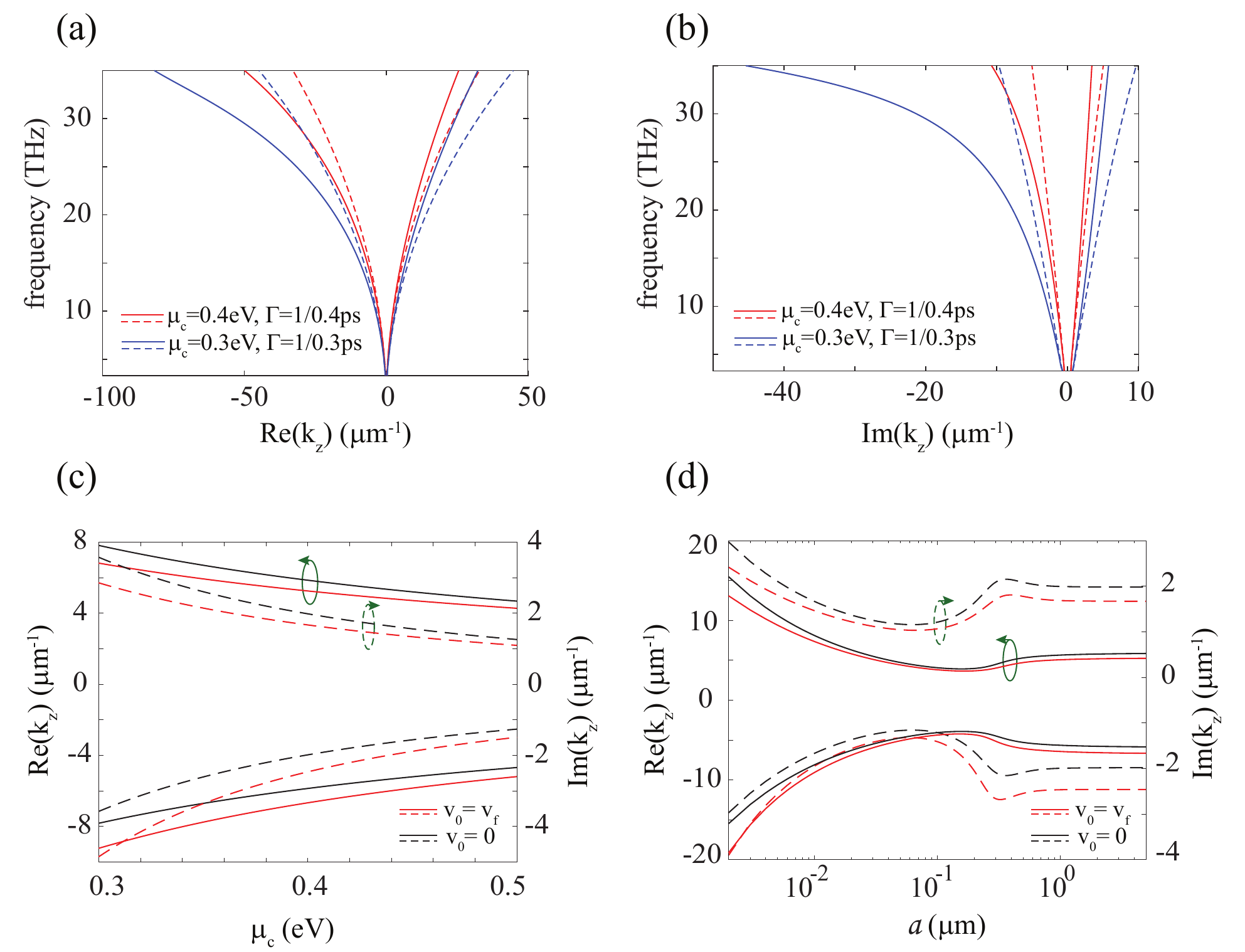}
\caption{(a) Real and (b) imaginary parts of wavenumber for biased graphene coating with various scattering rates, here we assume $k_z=Re(k_z)-i Im(k_z)$. Solid and dashed lines represent the biased ($v_\emph{0}=v_\emph{f}$) and unbiased ($v_\emph{0}=0$) cases, respectively. Real and imaginary part of wavenumber of the $SPP$ mode at $f=15$ $THz$ versus (c)  chemical potential and (d) the radius of the fiber for $v_\emph{0}=v_\emph{f}$ and $v_\emph{0}=0$.}\label{fig:loss}
\end{figure}
Fig. \ref{fig:geo}(b) and (c) depict the dispersion characteristics obtained from our analytical expression for biased (solid lines) and unbiased (dashed lines) cases. In this context, we assume a relative permittivity of $\varepsilon_{\emph{core}}=2.5$ for the lossless fiber, similar to polymers like Polystyrene \cite{sultanova2009dispersion}. Throughout this study, unless otherwise specified, the chemical potential, scattering rate, and temperature of graphene are assumed to be $\mu_\emph{c}=0.4$ $eV$, $\Gamma=1/(0.4$$ps$), and $T=300$ $K$, respectively. Red curves in Fig. \ref{fig:geo}(b) and (c) present the dispersion diagram of the $SPP$ mode in the fiber, which exhibits a pronounced nonreciprocal response (i.e., wavenumbers of these $SPP$ modes propagating in $+z$ and $-z$ directions are notably nonsymmetric). In panel (c) of Fig. \ref{fig:geo}, which is a magnified view of panel (b) near the origin, dispersion graphs of other guided modes such as $HE_{\emph{11}}$ and $TM_{\emph{10}}$ are also depicted. It is worth noting that the strength of nonreciprocity manifested in the term ($\tilde{\omega}=\omega - v_\emph{0} k_\emph{z}$) depends upon the relative values of $\omega$ and $v_\emph{0} k_\emph{z}$, which implies that in order to achieve a high level of nonreciprocity, these two values should be close to each other. As an illustration, consider the $HE_{\text{11}}$ mode at $12$ $THz$, in this case, the disparity between the real parts of the wave vectors of the two propagation directions is quantified by $\Delta k_\emph{z} = k_\emph{z}^- + k_\emph{z}^+ = -30.1442 \times 10^{-6}$ $\mu m^{-1}$ ( $k_\emph{z}^+$ for (right-going/$+z$) and $k_\emph{z} ^-$ for (left-going/$-z$)). Similarly, for the $TM_{\text{10}}$ mode at $25$ $THz$, the difference in $k_z$ is $\Delta k_z = -19.1925\times 10^{-6}$ $\mu m^{-1}$. These differences are much smaller than the values of $k_z$ in $HE_{\text{11}}$ and $TM_{\text{10}}$ modes. However, the difference in wave vectors for two propagation directions for the $SPP$ case is comparatively much larger. This explains why the $SPP$ mode demonstrates a robust nonreciprocal response, attributed to its higher value of $k_\emph{z}$ and strong field confinement near the graphene coating, in contrast to other modes, which remain practically unaffected.

We also present the field distribution of the $SPP$ guided mode at $f=15$ $THz$ for two distinct polarities of the DC electrical biasing with respect to the guided wave propagation along the fiber for two different radius of fiber in figure \ref{fig:geo}. Figure \ref{fig:geo}(d)-(g) present time snapshots of the real part of $z$-component of normalized electric field profile at $f=15$ $\mathrm{THz}$. In Fig. \ref{fig:geo} (d) and Fig. \ref{fig:geo}(e), radius of the fiber is assumed to be $a=5$ $\mu m$ and the field distributions are plotted for positive ($v_\emph{0}=v_\emph{f}$) and negative ($v_\emph{0}=-v_\emph{f}$) bias polarities, respectively. Similar figures have been demonstrated for $a=0.1$ $\mu m$ case in Fig. \ref{fig:geo}(f) and Fig. \ref{fig:geo}(g). These field distributions vividly demonstrate how the $SPP$ mode, concentrated around the graphene coating, has been "dragged" or "opposed" by the drift electrons.

In order to elucidate further the nonreciprocal response, we also investigate the influence of graphene loss on the real and imaginary parts of the guided wavenumber for the $SPP$ mode. Figures \ref{fig:loss} (a) and (b) show the dispersion characteristics of the structure while considering two different values for the electron scattering rate. As demonstrated in Fig. \ref{fig:loss}(a), the real part of the wavenumber tilts to the left when the graphene is biased, implying that by biasing the graphene coating, the guided $SPP$ propagating in the same direction as the drift electrons (right-going/$+z$), propagates with smaller wavenumber than the opposite direction (left-going /$-z$). Fig. \ref{fig:loss}(b) depicts imaginary parts of the wavenumber. This visualization highlights two key aspects: (1) in the presence of a nonzero electrical bias, the left-going/$-z$ mode experiences stronger attenuation than the right-going/$+z$ mode. This is due to the plasmons being dragged by the drift electrons in one direction and opposed in the opposite direction; (2) compared with the unbiased case, loss of the wave decreases in the same direction as the drift of electrons while increasing in the opposite direction.

Since the conductivity of graphene can be dynamically tuned using its chemical potential (via chemical doping or electric bias doping), here we explore the role of chemical potential in nonrecprocity of the $SPP$ mode. Figure \ref{fig:loss}(c) shows real and imaginary parts of the $SPP$'s wavenumber at $f=15$ $THz$ in the presence and absence of the electrical bias as a function of the chemical potential of the graphene coating. The results clearly demonstrate that tuning $\mu_\emph{c}$ allows us to tailor the nonreciprocal response. Hence, based on these findings, it can be inferred that the proposed structure holds potentials for serving as a tunable mid-IR nonreciprocal device.


In this work, we have studied the nonreciprocal response of an electrically biased graphene-coated fiber considering transverse magnetic, hybrid, and surface plasmon polariton modes. We have also presented the effects of the fiber radius, the chemical potential and material loss of graphene coating on the nonreciprocal response and how they are manifested in real and imaginary parts of modal wavenumber. The proposed structure may find possibilities in controlling and manipulating the excited $SPP$ modes inside of the graphene-coated optical fibers with potential applications in $\mathrm{THz}$ systems, microwave devices, and nanophotonics networks.

The authors express their thanks to Prof. M{\'a}rio G. Silveirinha of the University of Lisbon for valuable discussions.


This work is supported in part by the National Science Foundation (NSF) Emerging Frontiers in Research and Innovations (EFRI) Program grant $\#$1741693.

\nocite{*}


\end{document}